\documentclass[letter,longauth]{aa} 
\usepackage{graphicx}
\usepackage{txfonts}
\begin{document}
   \title{First detection of ND in the solar-mass protostar IRAS16293-2422\thanks{\emph{Herschel} is an ESA   space observatory with science instruments provided by
       European-led principal Investigator consortia and with
       important participation from NASA}}


    \author{
Bacmann A.\inst{1,2,30} \and
Caux E. \inst{6,29} \and
Hily-Blant P. \inst{1} \and
Parise B. \inst{10} \and
Pagani L. \inst{8} \and
Bottinelli S. \inst{6,29} \and
Maret S. \inst{1} \and
Vastel C. \inst{6,29} \and
Ceccarelli C. \inst{1,2,30} \and
Cernicharo J. \inst{9} \and
Henning T. \inst{26} \and
Castets A. \inst{1} \and
Coutens A. \inst{6,29} \and
Bergin E.A. \inst{25} \and
Blake G.A. \inst{3} \and
Crimier N. \inst{1,9} \and
Demyk K. \inst{6,29} \and
Dominik C. \inst{12,13} \and
Gerin M. \inst{28} \and
Hennebelle P. \inst{28} \and
Kahane C. \inst{1} \and
Klotz A. \inst{6,29} \and
Melnick G. \inst{18} \and
Schilke P. \inst{10,20} \and
Wakelam V. \inst{2,30} \and
Walters A. \inst{6,29} \and
Baudry A. \inst{2,30} \and
Bell T. \inst{3} \and
Benedettini M. \inst{4} \and
Boogert A. \inst{5} \and
Cabrit S. \inst{8} \and
Caselli P. \inst{7} \and
Codella C. \inst{11} \and
Comito C. \inst{10} \and
Encrenaz P. \inst{8} \and
Falgarone E. \inst{28} \and
Fuente A. \inst{14} \and
Goldsmith P.F. \inst{15} \and
Helmich F. \inst{16} \and
Herbst E. \inst{17} \and
Jacq T. \inst{2,30} \and
Kama M. \inst{12} \and
Langer W. \inst{15} \and
Lefloch B. \inst{1} \and
Lis D. \inst{3} \and
Lord S. \inst{5} \and
Lorenzani A. \inst{11} \and
Neufeld D. \inst{19} \and
Nisini B. \inst{24} \and
Pacheco S. \inst{1} \and
Pearson J. \inst{15} \and
Phillips T. \inst{3} \and
Salez M. \inst{8,28} \and
Saraceno P. \inst{4} \and
Schuster K. \inst{21} \and
Tielens X. \inst{22} \and
van der Tak F.F.S. \inst{16,27} \and
van der Wiel M.H.D. \inst{16,27} \and
Viti S. \inst{23} \and
Wyrowski F. \inst{10} \and
Yorke H. \inst{15} \and
Faure, A. \inst{1} \and
Benz, A. \inst{31} \and
Coeur-Joly, O. \inst{6,29} \and
Cros, A. \inst{6,29} \and
G\"usten, R. \inst{10} \and
Ravera, L. \inst{6,29} 
          }

   \institute{
Laboratoire d'Astrophysique de Grenoble, UMR 5571-CNRS, Universit\'e Joseph Fourier, Grenoble, France
\and Universit\'{e} de Bordeaux, Laboratoire d'Astrophysique de Bordeaux, Floirac, France
\and California Institute of Technology, Pasadena, USA
\and INAF - Istituto di Fisica dello Spazio Interplanetario, Roma, Italy
\and Infrared Processing and Analysis Center,  Caltech, Pasadena, USA
\and Centre d'Etude Spatiale des Rayonnements, Universit\'e Paul Sabatier, Toulouse, France
\and School of Physics and Astronomy, University of Leeds, Leeds UK
\and LERMA and UMR 8112 du CNRS, Observatoire de Paris,  Paris, France
\and Centro de Astrobiolog\'{\i}a, CSIC-INTA, Madrid, Spain
\and Max-Planck-Institut f\"{u}r Radioastronomie, Bonn, Germany
\and INAF Osservatorio Astrofisico di Arcetri, Florence Italy
\and Astronomical Institute 'Anton Pannekoek', University of Amsterdam, Amsterdam, The Netherlands
\and Department of Astrophysics/IMAPP, Radboud University Nijmegen,  Nijmegen, The Netherlands
\and IGN Observatorio Astron\'{o}mico Nacional, Alcal\'{a} de Henares, Spain
\and Jet Propulsion Laboratory,  Caltech, Pasadena, CA 91109, USA
\and SRON Netherlands Institute for Space Research, Groningen, The Netherlands
\and Ohio State University, Columbus, OH, USA
\and Harvard-Smithsonian Center for Astrophysics, Cambridge MA, USA
\and Johns Hopkins University, Baltimore MD,  USA
\and Physikalisches Institut, Universit\"{a}t zu K\"{o}ln, K\"{o}ln, Germany
\and Institut de RadioAstronomie Millim\'etrique, Grenoble - France
\and Leiden Observatory, Leiden University, Leiden, The Netherlands
\and Department of Physics and Astronomy, University College London, London, UK
\and INAF - Osservatorio Astronomico di Roma, Monte Porzio Catone, Italy
\and Department of Astronomy, University of Michigan, Ann Arbor, USA
\and Max-Planck-Institut f\"ur Astronomie, Heidelberg, Germany
\and Kapteyn Astronomical Institute, University of Groningen, The Netherlands
\and Laboratoire d'Etudes du Rayonnement et de la Mati\`ere en Astrophysique, UMR 8112  CNRS/INSU, OP, ENS, UPMC, UCP, Paris, France
\and CNRS/INSU, UMR 5187, Toulouse, France
\and CNRS/INSU, UMR 5804, Floirac cedex, France
\and Institute of Astronomy, ETH Z\"urich, Z\"urich, Switzerland
          }

   \date{Received 31 May 2010 / Accepted 30 June 2010}

 
  \abstract
   {In the past decade, much progress has been made  in characterising the processes leading to the enhanced deuterium fractionation observed in the ISM and in particular in the cold, dense parts of star forming regions such as protostellar envelopes. Very high molecular D/H ratios have been found for saturated molecules and ions. However, little is known about the deuterium fractionation in radicals, even though simple radicals often represent an intermediate stage in the formation of more complex, saturated molecules. The imidogen radical NH is such an intermediate species  for the ammonia synthesis  in the gas phase. Many of these light molecules however have their fundamental transitions in the submillimetre domain and their detection is hampered by the opacity of the atmosphere at these wavelengths. \emph{Herschel}/HIFI represents a unique opportunity to study the deuteration and formation mechanisms of species not observable from the ground.}
   {We searched here for the deuterated radical ND in order to determine the deuterium fractionation of imidogen and constrain the deuteration mechanism of this species.} 
   {We observed the solar-mass  Class 0 protostar IRAS16293-2422 with the heterodyne instrument HIFI in Bands 1a (480 -- 560\,GHz), 3b (858 -- 961\,GHz), and 4a (949 -- 1061\,GHz) as part of the \emph {Herschel} key programme CHESS (Chemical \emph{HErschel} Surveys of Star forming regions).}
   {The deuterated form of the imidogen radical ND was detected  and securely identified with 2 hyperfine component groups of its fundamental transition (N=0 --1) at 522.1 and 546.2\,GHz, in absorption against  the continuum background emitted from the nascent protostar. The 3 groups of hyperfine components of its hydrogenated  counterpart NH were also detected in absorption. The absorption arises from the cold envelope, where many deuterated species have been shown to be abundant. The estimated column densities are $\sim 2\times 10^{14}$\, cm$^{-2}$ for NH and $\sim 1.3\times 10^{14}$\,cm$^{-2}$ for ND. We derive a very high deuterium fractionation with an [ND]/[NH] ratio of between 30 and 100\%.}
   {The deuterium fractionation of imidogen is of the same order of magnitude as that in other molecules, which suggests that an efficient deuterium fractionation mechanism is at play. We discuss two possible formation pathways for ND, by means of either the reaction of N$^+$ with HD, or deuteron/proton exchange with NH. }


   \maketitle
%

\section{Introduction}

The envelopes of low-mass Class 0 protostars have been known to be characterised by high levels of molecular deuteration, as shown e.g., by Loinard et al. (\cite{loinard02}), Roberts et al. (\cite{roberts07}, \cite{roberts02}),  and the detection of multiply deuterated species (Ceccarelli et al. \cite{cc1998} for D$_2$CO, Parise et al. \cite{parise02} for CHD$_2$OH, Parise et al. \cite{parise04} for   CD$_3$OH, van der Tak et al. \cite{vandertak} for ND$_3$). The high abundance of deuterated molecules is believed to originate in the  pre-stellar phase (Bacmann et al. \cite{bacmann}):  at the low temperatures ($< 20$\,K)  and high densities ($n > 10^{4}$ cm$^{3}$) prevailing in pre-stellar cores (and similarly in protostellar envelopes), heavy molecules such as CO are depleted on the dust grains, so that species such as H$_2$D$^+$, D$_2$H$^+$, or D$_3^+$ reach high abundances. The increase in the [H$_2$D$^+$]/[H$_3^+$]  ratio (and similar ratios involving multiply deuterated counterparts) enhances both the molecular deuterium fractionation by means of gas-phase reactions and the atomic D/H ratio, therefore also increasing the deuteration of species forming on grain surfaces (e.g. Roberts et al. \cite{roberts03}).

Most deuterated molecules detected to date are either saturated molecules or ions. The deuteration of radicals, which are important intermediate pieces leading to the formation of saturated molecules, has so far not been thoroughly investigated. Their observation could however provide an important clue to the formation process of the major deuterated species. For example, it is not yet completely understood if saturated deuterated molecules are formed directly (from deuterated ion intermediates, whose dissociative recombination may lead to deuterated radicals) like their hydrogenated counterparts, or if they are formed by proton-deuteron exchange starting from the saturated hydrogenated molecule. The observation of the D/H ratio in the radical may help us to disentangle these two processes. With this in mind, we study the deuteration ratio in the NH radical, which is formed from the dissociative recombination of NH$_2^+$ and NH$_3^+$, two precursors of ammonia in the gas-phase. Roueff et al. (\cite{roueff}) argued that highly-deuterated ammonia can form from pure gas phase reactions.  These models however require highly deuterated radicals, and we propose to observationally test this model.

Light hydrides such as NH have fundamental transitions in the submillimetre domain: NH possesses 3 groups of hyperfine transitions around 950\,GHz--1\,THz  and its deuterated counterpart ND has its 3 groups of fundamental hyperfine structure transitions around 490--550\,GHz. At all of these frequencies apart from 490\,GHz, the atmosphere is opaque and precludes observations from the ground. However, these transitions can all be observed with the HIFI instrument on board the \emph{Herschel} Space Observatory. Since hydrides generally have high Einstein $A_{\rm ul}$ coefficients, only the ground state is expected to be significantly populated. As a consequence, it is possible to detect these species in absorption against the continuum background originating from the heating of the nascent protostar. Moreover, as already mentioned above, deuterated molecules are very abundant in cold, dense medium, so that the envelopes of young protostars represent ideal targets to study the deuteration of the NH radical.

In the course of the CHESS key programme (Ceccarelli et al. \cite{overview}), we carried out a spectral survey of the frequencies covered by HIFI in the young low-mass protostar IRAS\,16293-2422 (hereafter IRAS16293). The frequency bands contain the fundamental hyperfine transitions of ND and NH. In this Letter, we report  the first detection by HIFI of ND, as well as the detection of the hyperfine transitions of NH. Section 2 describes the observations and the determination of the radicals' column densities, and Sect. 3 discusses the implications of the derived deuterium fractionation on our understanding of deuteration processes in N-bearing species.

\section{Observations and results}

The solar-mass protostar IRAS16293 was observed with the HIFI instrument (de Grauuw et al. \cite{degrauuw}; Roelfsema et al. \cite{roelfsema}) on board the \emph{Herschel} Space Observatory (Pilbratt et al., \cite{pilbratt}), as part of the HIFI guaranteed time key programme CHESS (Ceccarelli et al. \cite{overview}). A full spectral coverage of bands 1a (480 -- 560\,GHz), 3b (858 -- 961\,GHz), and 4a (949 -- 1061\,GHz) was performed on 2010 March 1, 19, and 3, respectively, using the HIFI spectral scan double beam switch (DBS) mode with optimisation of  the continuum. In this mode, the HIFI acousto-optic wide band spectrometer (WBS) was used, providing a spectral resolution of 1.1\,MHz ($\sim$0.6 km s$^{-1}$ at 520 GHz and 0.3 km s$^{-1}$ at 1 THz) over an instantaneous bandwidth of 4x1\,GHz. The targeted coordinates were $\alpha_{2000}$ = 16$^h$ 32$^m$ 22$\fs$75, $\delta_{2000}$ = $-$ 24$\degr$ 28$\arcmin$ 34.2$\arcsec$. The DBS reference positions were situated approximately 3\arcmin\,east and west of the source. The beam sizes at the frequencies of the ND and NH transitions are about 41\arcsec\,and 22\arcsec, respectively, and the theoretical main beam and forward efficiencies are about 0.72 and 0.96, respectively. Calibration uncertainties are $\leq 16$\,\% for band 1 and $\leq 32$\,\% for bands 3 and 4 (June 2010 estimates).

The data were processed using the standard HIFI pipeline up to frequency and amplitude calibrations (level 2) with the ESA-supported package HIPE 2.8 (Ott et al. \cite{ott}). A single local oscillator tuning spectrum consists, for each polarisation, of 4 sub-bands of $\sim\,1$\,GHz for the SIS bands (1 to 5). The 1\,GHz scans are then exported as FITS files into CLASS/GILDAS  format\footnote{http://www.iram.fr/IRAMFR/GILDAS} prior to data reduction and analysis using generic spectral survey tools developed in CLASS by our group. Spurious features in the spectra were first removed in each 1\,GHz scan, and a low order (typically 3) polynomial baseline was then fitted over line-free regions to correct for residual bandpass effects. These polynomials  were subtracted and used to determine an accurate continuum level by calculating their medians. Higher order  oscillations in the baseline are possible, but their amplitudes remain low in the spectra considered here. The continuum values obtained in this way are well fitted by a degree 1 polynomial over the frequency range $\sim$500--1200\,GHz. Sideband deconvolution is computed with the minimisation algorithm of Comito \& Schilke (\cite{comito}) implemented into CLASS90 using the baseline-subtracted spectra. The single side-band continuum derived from the polynomial fit at the considered frequency (Table\,\ref{conti}) was eventually added to the spectra. The temperatures were converted to the $T_{\rm mb}$ scale, using the theoretical values of the main beam and forward efficiencies given above.

\begin{figure}
   \centering
   \includegraphics[width=7.8cm]{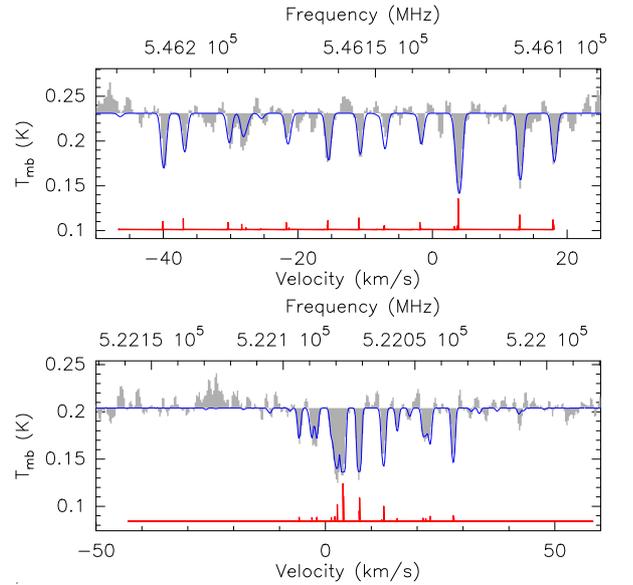}
   \caption{ND ($N_J=0_1-1_1$) transition at 546.15\,GHz (upper panel) and ND ($N_J=0_1 - 1_2$) transition at
   522.1\,GHz (lower panel). The filled histograms show the observed spectra and the solid line is the fit of the hyperfine structure performed using the CLASS software. At the bottom of each panel is a sketch of the positions and relative intensities of the hyperfine components. Components at the same frequency have been slightly put apart on the velocity axis for visualization reasons. The origin of the velocity axis was chosen so that the strongest component is at the LSR velocity of the source, $\varv=3.8$\,km\,s$^{-1}$. The emission feature towards 522.12\,GHz coincides with a methanol line.}
              \label{ND}%
    \end{figure}
\begin{figure}
\centering
\includegraphics[width=7.8cm]{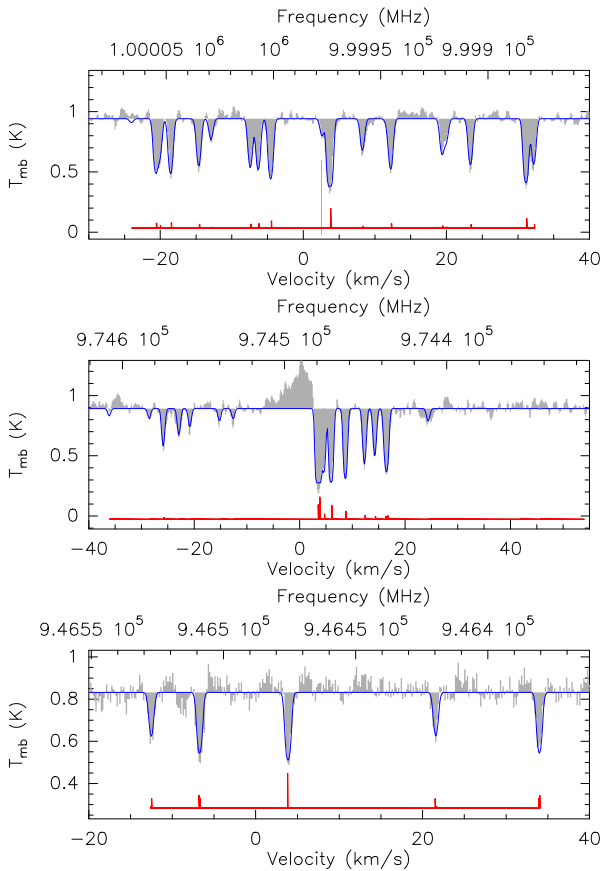}
\caption{NH ($N_J=0_1 -1_0$) transition at 946.5\,GHz (top panel), NH ($N_J=0_1 - 1_2$) transition at 974.5\,GHz (middle panel) and NH ($N_J=0_1 - 1_1$) transition at 1000.0\,GHz (bottom panel). The solid line fitting the spectra and the sketch at the bottom of each panel have the same signification as in Fig.\,\ref{ND}. Note the HCN line towards 974.48\,GHz.}
\label{NH}
\end{figure}

We detect for the first time the 3 groups of hyperfine transitions of ND at 491.9, 522.1, and 546.2\,GHz of the N{\bf =} 0  -- 1 transition  as well as the 3 groups of hyperfine components of the NH fundamental transitions at 946.5, 974.5\,GHz, and 1\,THz (see Klaus et al. \cite{klaus} for an energy diagram of the N= 0--1 transition of NH). The transitions are seen in absorption against the continuum from the protostar and the hyperfine structure is partially resolved. The transitions of ND at 522.1 and at 546.2\,GHz are presented in Fig.\,\ref{ND}. 
From Fig.\,\ref{ND}, the lines can be unambiguously attributed to ND (but see Olofsson et al. \cite{olofsson}, who tentatively assign an unidentified line to ND). The transition of ND at 491.9\,GHz is not shown here because some of its components are strongly blended with a bright H$_2$CO line at 491.9683\,GHz, with an SO$_2$ line at 491.9347 GHz and with an HDCO line at 491.9370 GHz. This transition was not used in our analysis and will not be discussed further. Figure\,\ref{NH} presents the 3 detected hyperfine groups of the fundamental  transition for NH.


The analysis was carried out by fitting the hyperfine structure using the HFS method in the CLASS software\footnote{see the CLASS manual at http://www.iram.es/IRAMES/\-otherDocuments/manuals/\-index.html for more details}. The line parameters (frequencies, Einstein $A_{\rm ul} $ coefficients, and level degeneracies), originally from spectroscopic studies of Klaus et al. (\cite{klaus}) and Saito \& Goto (\cite{saito}) for NH and ND, respectively, were taken from the Cologne Database for Molecular Spectroscopy (M\"uller et al. \cite{cdms1}, \cite{cdms2}). The routine assumes a zero-baseline spectrum (i.e., from which the continuum emission has been subtracted) and therefore fits a line profile of the form $(1-e^{-\tau})(J_{\nu}(T_{\rm ex})-J_{\nu}(T_{\rm bg})- T_{\rm c}$), where $\tau$ is the optical depth, $T_{\rm c}$ the intensity of the continuum, and $J_{\nu}(T_{\rm ex})$ and $J_{\nu}(T_{\rm bg})$ the radiation temperature for the excitation temperature $T_{\rm ex}$ and the radiation temperature of the cosmological background, respectively. The parameters given by the HFS fitting procedure are reported in Table\,{\ref{coldens}}, Cols. (2) to (4). From the total optical depth $\tau$ given by the fit, we can infer the opacities of single hyperfine components by multiplying $\tau$ by the relative intensities of these components. The total column density of the considered species is  given by
\begin{equation}
\label{equation}
N=\frac{8\pi\nu^3}{c^3}\frac{Q(T_{\rm ex})}{g_{\rm up} A_{\rm ul}}\frac{e^\frac{E_{\rm up}}{kT_{\rm ex}}}{e^\frac{h\nu}{k T_{\rm ex}}-1}\int{\tau d\varv}
\end{equation}
where $\nu$ is the frequency of the transition, $c$ the velocity of light, $g_{\rm up}$ the upper level degeneracy, $A_{\rm ul}$ the spontaneous emission Einstein coefficient, $T_{\rm ex}$ the excitation temperature, $Q(T_{\rm ex})$ the partition function at $T_{\rm ex}$, and $E_{\rm up}$ the upper level energy. For the values of the excitation temperature and the frequencies considered here, the factor with the exponentials is nearly equal to 1. The velocity integrated optical depth of component $i$ is given by $\tau_{\rm i}\Delta \varv\frac{\sqrt{\pi}}{2\sqrt{\ln{2}}}$ with $\tau_{\rm i}$ the optical depth of component $i$ and $\Delta \varv$ the linewidth determined by the HFS method (full width at half maximum). The obtained values of the column density are given for each group of hyperfine components in Col. (6) of Table\,\ref{coldens}.
In our derivation of the excitation temperature, it was assumed that the zone where the absorption arises completely fills the emission zone of the continuum. 
In the HFS fitting method, the determination of the optical depth comes only from the measured relative ratios of the hyperfine components and is independent of any assumption about the filling factor. Overall, since the exponential factor is close to 1 in Eq.\ref{equation}, and the partition function depends only mildly on $T_{\rm ex}$ (in the case of NH for example, it increases by 10\% when $T_{\rm ex}$ increases from 10 to 15\,K), the inferred value of the column density depends very little on the assumptions about the value of the continuum, beam efficiency values, or absolute calibration uncertainties.

The column density found for NH is $N({\rm NH})=(2.0 \pm 0.8)\times 10^{14}$\,cm$^{-2}$. For ND, the column density values found  from the 2 observed transitions differ slightly but are consistent within the error bars. A reasonable estimate for the ND column density is therefore: $N({\rm ND})= (1.3 \pm 0.8)\times10^{14}$\,cm$^{-2}$.

With these values, we find a  imidogen deuterium fraction of [ND]/[NH] $\sim$ 30--100\% in the envelope of IRAS16293. We note however that because of the differences in beam sizes, the region sampled by the ND observations is larger and on average characterised by lower densities and temperatures than the region sampled by the NH observations for which the contribution of the warmer inner regions is larger. To take these effects into account, a  more detailed radiative transfer modelling -- taking into account the source structure and the exact coupling between the source and the instrument -- would be necessary. The similarity of the determined linewidths (around 0.5\,km s$^{-1}$ in both lines, when deconvolved from the resolution of the spectrometer) is consistent with the bulk of the signal from both molecules arising from the same (cold) region.

\begin{table}
\caption{Measured rms (at a frequency resolution of 1.1\,MHz) and assumed continuum values  for the different  transitions  (in $T_{\rm mb}$ scale). }             
\label{conti}      
\centering                          
\begin{tabular}{c c c c c}        
\hline\hline                 
HIFI band  &  Transition & Frequency  & rms & Continuum \\    
 & ($N _J$ -- $N\arcmin_J\arcmin$) & (GHz) & (mK) & (mK)\\
\hline                        
1a & ND  (0$_1$--1$_2$) & 522.1 &  12 & 203 $\pm$ 22\\      
1a & ND  (0$_1$--1$_1$) & 546.2 &  15 & 230 $\pm$ 24\\
3b & NH  (0$_1$--1$_0$) & 946.5 &  58 & 832 $\pm$ 80\\
4a & NH  (0$_1$--1$_2$) & 974.5 & 49 & 892 $\pm$ 143\\
4a & NH  (0$_1$--1$_1$) & 1000.0 & 50 & 942 $\pm$ 146\\ 
\hline                                   
\end{tabular}
\end{table}

\begin{table*}
\caption{Parameters given by the HFS fit of each ND and NH transition.  $\tau$ is the total optical depth, $T_{\rm ex}$ the excitation temperature, $J_{\nu}(T_{\rm ex})$ the radiation temperature at the frequency of the transition, $J_\nu(T_{\rm bg})$ the radiation temperature of the cosmological background at $T_{\rm bg}=2.7$\,K, $T_{\rm c}$ the brightness temperature of the continuum,  $\Delta \varv$ the linewidth (full width at half maximum), and $N$ the total measured column density in the given species. $T_{\rm ex}$ was deduced from Col. (2). Because of uncertainties in some stages of the data reduction process, we estimate that the linewidths are not known to better than about 0.25 MHz. It is this uncertainty that we quote here in Col. (3).}
\label{coldens}
\centering
\begin{tabular}{c c c c c c}
\hline\hline
Molecule  & $\tau(J_{\nu}(T_{\rm ex}) - J_\nu(T_{\rm bg})-T_{c})$ & $\Delta \varv$ & $\tau$ & $T_{\rm ex}$ & $N$ \\ 
\& transition& K & km s$^{-1}$ & & K & cm$^{-2}$\\
\hline
ND (0$_1$--1$_2$) & $-$2.57 $\pm$ 0.33 & 0.74 $\pm$ 0.14 & 28.8 $\pm$ 3.9 & 4.6 & $(1.7 \pm 0.6) \times 10^{14}$\\
ND (0$_1$--1$_1$) & $-$1.18 $\pm$ 0.13 & 0.82 $\pm$ 0.14 & 7.9 $\pm$ 2.1 & 4.5 & $(0.9 \pm 0.4) \times 10^{14}$\\
NH (0$_1$--1$_0$) & $-$3.78 $\pm$ 0.58 & 0.60 $\pm$ 0.08 & 8.4 $\pm$ 2.1 & 9.5 &  $(2.0 \pm 0.8) \times 10^{14}$\\
NH (0$_1$--1$_2$) & $-$27.3 $\pm$ 0.27 & 0.64 $\pm$ 0.08 & 34.9 $\pm$ 4.8 & 7.7 & $(1.8 \pm 0.4) \times 10^{14}$\\
NH (0$_1$--1$_1$) & $-$18.6 $\pm$ 0.75 & 0.63 $\pm$ 0.08 & 25.1 $\pm$ 1.6 & 8.8 & $(2.2 \pm 0.4) \times 10^{14}$\\
\hline
\end{tabular}
\end{table*}

%
%

\section{Discussion and conclusions}

The very high deuterium fractionation observed in IRAS16293 is of the same order of magnitude as the molecular D/H ratio previously measured in other species  in the envelope of IRAS16293, i.e., [HDCO/H$_2$CO]=15\%, [NH$_2$D]/[NH$_3$] = 10\% (van Dishoeck et al. \cite{vandishoeck}) [CH$_2$DOH]/[CH$_3$OH]=30\% (Parise et al. \cite{parise04}). The [ND]/[NH] ratio even appears to be the highest measured deuterium fractionation in this source, though this result needs to be confirmed with a more detailed modelling. 


Our measured ratio is consistent with the gas-phase predictions of Roueff et al. (\cite{roueff}), although this model assumes that NH is a major product of  the dissociative recombination of N$_2$H$^+$, a result that has since been refuted (Molek et al. \cite {molek}; Talbi \cite{talbi}). 

The main path for the gas-phase formation of NH in the dense interstellar medium is believed to be initiated by the endothermic (85\,K) reaction
N$^+$ + H$_2$ $\rightarrow$ NH$^+$ + H (e.g., Galloway \& Herbst \cite{galloway}), which turns into NH$_2^+$, NH$_3^+$, and NH$_4^+$ after successive additions of H$_2$. We then expect NH to be mostly a product of the dissociative recombination of electrons with  NH$_2^+$ and NH$_3^+$ (Thomas et al. \cite{thomas}), while NH$_4^+$ leads to ammonia. 
One possible way of forming ND is by means of the reaction N$^+$ + HD $\rightarrow$ ND$^+$ + H. This reaction is less endothermic than the corresponding reaction with H$_{2}$ (16\,K -- Marquette et al. \cite{marquette}) so is more likely to occur at low temperatures. Furthermore, experiments indicate that the dissociative recombination of deuterated ions tends to eject H atoms preferentially (see Roueff et al. \cite{roueff} and references therein). The dissociative recombination of e.g. NHD$^+$ or NH$_2$D$^+$ should therefore favour the production of ND with respect to NH. On the other hand, the low abundance of HD ($\sim 3.2\times 10^{-5}$ [H$_2$], Linsky \cite{linsky}) might make it difficult to attain [ND]/[NH] ratios around our measured value.

An alternative means of forming ND is by proton-deuteron reactions with NH, i.e., NH + H$_2$D$^+$ $\rightarrow$ NHD$^+$ + H$_2$, followed by a dissociative recombination of NHD$^+$, which leads predominantly to the deuterated molecule. 
This scheme is similar to that proposed by Rodgers \& Charnley (\cite{rodgers}) to account for the high abundances of doubly deuterated ammonia, 
and relies on there being high abundances of species such as H$_2$D$^+$ . In cold dense regions depleted of heavy elements such as CO, the ratio of [H$_2$D$^+$]/[H$_3^+$] is indeed highly enhanced (e.g. Caselli et al. \cite{caselli}, Caselli et al. \cite{caselli03}, Roberts et al. \cite{roberts03}). 
The ion H$_2$D$^+$ has indeed been detected in the cold envelope of IRAS16293 by Stark et al. (\cite{stark}), who estimated its abundance to be $2\times10^{-9}$.
Although both routes of ND formation considered here seem plausible, it is not possible for us to conclude anything about their relevance without a more complete chemical model.

We have reported the first detection of the radical ND by the high resolution heterodyne instrument HIFI on board the \emph{Herschel} Space observatory, towards the young solar-mass protostar IRAS16293. The hyperfine structure is seen in absorption against the continuum background from the protostar, for both ND and NH, which implies that the absorption arises from the cold envelope around the hot corino. The deuterium fractionation measured is very high, with a ratio [ND]/[NH] between 30 and 100\%. Detailed radiative transfer and chemical modelling are needed to constrain the deuteration and formation of NH  and this will be the subject of a future study.

\begin{acknowledgements}
  HIFI has been designed and built by a consortium of institutes and
  university departments from across Europe, Canada and the United
  States under the leadership of SRON Netherlands Institute for Space
  Research, Groningen, The Netherlands and with major contributions
  from Germany, France and the US. Consortium members are: Canada:
  CSA, U.Waterloo; France: CESR, LAB, LERMA, IRAM; Germany: KOSMA,
  MPIfR, MPS; Ireland, NUI Maynooth; Italy: ASI, IFSI-INAF,
  Osservatorio Astrofisico di Arcetri-INAF; Netherlands: SRON, TUD;
  Poland: CAMK, CBK; Spain: Observatorio Astron\'omico Nacional (IGN),
  Centro de Astrobiolog\'{\i}a (CSIC-INTA). Sweden: Chalmers
  University of Technology - MC2, RSS \& GARD; Onsala Space
  Observatory; Swedish National Space Board, Stockholm University -
  Stockholm Observatory; Switzerland: ETH Zurich, FHNW; USA: Caltech,
  JPL, NHSC.  We thank many funding agencies for financial support.
\end{acknowledgements}


\begin{thebibliography}{}


    \bibitem[2003]{bacmann}Bacmann, A., Lefloch, B., Ceccarelli, C., et al. 2003
    ApJ, 585, L55
    
    \bibitem[2008]{caselli}Caselli, P., Vastel, C., Ceccarelli, C., et al. 2008,
    A\&A, 492, 703
    
    \bibitem[2003]{caselli03}Caselli, P., van der Tak, F.F.S., Ceccarelli, C.,  Bacmann, A. 2003,
    A\&A, 403, 37
    
    \bibitem[2010]{overview}Ceccarelli, C., Bacmann, A., Boogert, A., et al. 2010, this volume

    \bibitem[1998]{cc1998}Ceccarelli, C., Castets, A., Loinard, L., et al., 1998,
    A\&A, 338, 43
    
    \bibitem[2002]{comito}Comito, C., \& Schilke, P., 2002, 
    A\&A, 395, 357
    
    \bibitem[2010]{degrauuw}de Graauw, T., , Helmich F.P., Phillips T.G., et al.  2010, 
    A\&A, 518, L6 
        \bibitem[1989]{galloway}Galloway, E.T., \& Herbst, E. 1989,
    A\&A, 211, 418
    
    
    
   \bibitem[1997]{klaus} Klaus, Th., Takano, S., \& Winnewisser, G. 1997,
  A\&A, 322, L1

   \bibitem[2007]{linsky}Linsky, J.L. 2007,
  Space Sci. Rev., 130, 367

   \bibitem[2002]{loinard02}Loinard, L., Castets, A., Ceccarelli, C., et al., 2002, 
   P\&SS, 50, 1205

    
    \bibitem[1988]{marquette}Marquette, J.-B., Rebrion, C., \& Rowe, B.R. 1988
    J. Chem. Phys, 89, 2041
    
    \bibitem[2007]{molek}Molek, C.D., McLain, J.L., Poterya, V., Adams, N.G., 2007, 
    J. Phys. Chem A, 111, 6760
    
   \bibitem[2005]{cdms2}M\"uller, H.S.P., Schl\"oder, F., Stutzki, J., \& Winnewisser, G., 2005, 
  J. Mol. Struct., 742, 215
  
   \bibitem[2001]{cdms1}M\"uller, H.S.P., Thorwirth, S. , Roth, D.A., \& Winnewisser, G. 2001, 
  A\&A, 370, L49
  
  \bibitem[2007]{olofsson}Olofsson, A.O.H., Persson, C.M., Koning, N., et al. 2007, A\&A, 476, 791
  
   \bibitem[2010]{ott}Ott, S. 2010, in ASP Conference Series, Astronomical Data Analysis Software and Systems XIX, Y. Mizumoto, K.-I. Morita, and M. Ohishi, eds., in press

  
    \bibitem[2002]{parise02}Parise, B., Ceccarelli, C., Tielens, A.G.G.M., et al. 2002,
    A\&A, 393, 49
    
    \bibitem[2004]{parise04}Parise, B., Castets, A., Herbst, E., et al. 2004,
    A\&A, 416, 159
    
    
    \bibitem[2010]{pilbratt}Pilbratt, G., Riedinger, J.R., Passvogel, T., et al. 2010, A\&A, 518, L1
    
    \bibitem[2002]{phillips}Phillips, T.G. and Vastel, C. 2003, in Chemistry as a Diagnostic of Star Formation, ed. C.L. Curry \& M. Fish, astro-ph/0211610
    
    
    \bibitem[2010]{roelfsema}Roelfsema, P.R., Helmich, F.P., Teyssier, D., et al. 2010,
    submitted
    
    \bibitem[2007]{roberts07}Roberts, H., Fuller, G.A., Millar, T.J., et al. 2007,
    A\&A, 381, 1026

   \bibitem[2003]{roberts03}Roberts, H., Herbst, E., \& Millar, T.J. 2003, 591, L41

   \bibitem[2002]{roberts02}Roberts, H., Fuller, G.A., Millar, T.J., et al. 2002,
   A\&A, 381, 1026
   
   \bibitem[2001]{rodgers}Rodgers, S.D. \& Charnley, S.B. 2001,
   ApJ, 553, 613
   
   \bibitem[2005]{roueff}Roueff, E., Lis, D.C., van der Tak, F.F.S., et al. 2005, 
   A\&A, 438, 585

    \bibitem[1993]{saito} Saito, S., \& Goto, M. 1993,
    ApJ, 410, L53
       \bibitem[2004]{stark}Stark, R., Sandell, G., Beck, S.C., et al. 2004, ApJ, 608, 341

 
    \bibitem[2009]{talbi}Talbi, D. 2009,
     Journal of Physics: Conference series, 192, 012015
    
    \bibitem[2005]{thomas}Thomas, R.D., Hellberg, F., Neau, A., et al. 2005, 
    Phys. Rev. A, 71, 032711
    
    
    \bibitem[2002]{vandertak} van der Tak, F.F.S., Schilke, P., M\"uller, H.S.P., et al. 2002,
    A\&A, 388, L53
 
     
   \bibitem[1995] {vandishoeck}van Dishoeck, E.F., Blake, G.A., Jansen, D.J., \& Groesbeck, T.D. 1995, 
  ApJ, 477, 760 
   
   


\end{thebibliography}
\end{document}